\documentclass[twocolumn,showpacs,preprintnumbers,superscriptaddress,prl]{revtex4}
\usepackage{subfigure}
\usepackage{graphicx,dcolumn,bm,color,latexsym,amssymb}
\usepackage[latin1]{inputenc}

\definecolor{Blue}{rgb}{0.0,0.0,1}
\definecolor{Red}{rgb}{1,0.0,0.0}
\definecolor{Green}{rgb}{0,0.5,0.0}

\begin{document}

\title{Entanglement and Bell's inequality violation above room temperature
in metal carboxilates.}

\author{A.M. Souza}
\author{D.O. Soares-Pinto}
\author{R.S. Sarthour}
\author{I.S. Oliveira}
\affiliation{Centro Brasileiro de Pesquisas F�sicas,
Rua Dr. Xavier Sigaud 150, Urca, Rio de Janeiro-RJ 22290-180,
Brasil.}
\author{M.S. Reis}
\author{P. Brand�o}
\affiliation{CICECO, Universidade de Aveiro, 3810-193
Aveiro,Portugal}
\author{A. M. dos Santos}
\affiliation{NSSD-MSTD, Oak Ridge National Laboratory, Oak Ridge-TN 37831-6475 USA.}

\date{\today}

\begin{abstract}
In the present work we show that a special family of
materials, the metal carboxylates, may have entangled states up to very high temperatures. From magnetic susceptibility measurements, we have estimated the
critical temperature below which entanglement exists in the cooper carboxylate  \{Cu$_2$(O$_2$CH)$_4$\}\{Cu(O$_2$CH)$_2$(2-methylpyridine)$_2$\}, and we have found this to be above room temperature ($T_e  \sim 630$ K). Furthermore, the results show that
the system remains maximally entangled until close to $\sim 100$ K and the Bell's
inequality is violated up to nearly room temperature ($\sim 290$ K).
\end{abstract}

\pacs{ }

\maketitle

\section{introduction}

Entanglement is the key resource for the majority of applications of the recent growing field of quantum information and quantum computation \cite{nielsenB,oliveira}. This unique
quantum phenomenon was until a few years ago thought to exist only in systems with small number of
particles at very low temperatures. However, recently it has been discovered
that entanglement can also be present in systems
containing a large number of particles at finite temperatures \cite{vedralprl,vedralrmp,vedralepl}. The presence of entangled states in thermal systems has been studied in a few experiments involving magnetic 
materials \cite{ghosh,brukner,rappoport,vertesi,souza,pinto}.

Vlatko Vedral \cite{vedral} has stated three basic motivations for
studying entanglement in many body systems: (i) the need to know the limits of the entanglement, i.e., how large systems
can support entanglement and how robust entanglement can be against temperature; (ii) the question
whether entanglement can be used as an order parameter for quantum phase transition and, (iii),
the need for novel materials which can be used for practical applications in quantum
computation and quantum communication. Materials in which robust and useful entangled state can be
found naturally could be of great relevance to design quantum solid state devices or as a source of entanglement \cite{chiara06}. By useful entanglement we mean entangled
states that can be used to implement quantum protocols, which are more efficient than their classical counterpart. For instance, in quantum cryptography applications and quantum communication complexity tasks, the useful entangled states are those who violate  Bell's inequalities \cite{brukner2,acin}.

Molecular magnets \cite{bogani} can be an excellent physical 
realization of spin chains, providing good opportunities for 
studying the above topics. In this class of materials, the intermolecular magnetic 
interactions are extremely weak compared to those within individual
molecules. A bulk sample, comprised by 
a set of non-interacting molecular clusters, is therefore completely described in
terms of independent clusters. From a physical point of view, a molecular magnet can  
combine classical properties found in any macroscopic magnet \cite{bogani} and quantum 
properties, such as quantum interference \cite{ramsay} and 
entanglement \cite{brukner,rappoport,vertesi,souza,pinto}. Recently, molecular magnets 
have been pointed out as good systems to be used in high-density information memories and also, due to their
long coherence times \cite{ardavan}, in spin based quantum 
computing devices \cite{affronte,troiani1,troiani2,michael,lehmann}. 

The existence of entangled states in molecular magnets is due to the fact that some molecular 
spin chains can have an entangled ground state. The separation between the ground state 
and the excited states energies is an important parameter which determine the 
temperature of entanglement ($T_e$), i.e. the temperature where the thermal state of the molecular magnet 
become separable. In the most simple chain, dimers, one can state that stronger is the exchange interaction energy, higher will be $T_e$ \cite{pinto}.

In this paper we show that a special family of molecular magnets (metal
carboxylates) can support entanglement above room temperature. We have found in the compound
\{Cu$_2$(O$_2$CH)$_4$\}\{Cu(O$_2$CH)$_2$(2-methylpyridine)$_2$\} that $T_e  \sim 630$ K. Furthermore, we could 
also conclude that the system remains in a pure
maximally entangled state up to $\sim 100$ K and the Bell's inequality can
be violated up to the room temperature ($ \sim 290$ K). This paper is organized as follows: In the next 
section we give a brief description of the system studied here. The following section contains a study of the 
entaglement in the compound and in the last section, some commets and conclusions are drawn.

\section{The compound \label{sec2}}

\begin{figure}[tbp]
\begin{center}
\includegraphics[width=7.0cm]{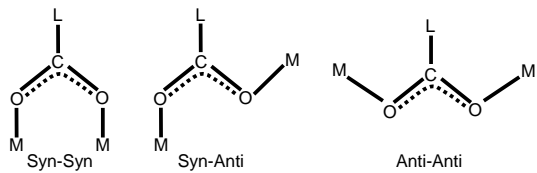}
\end{center}
\caption{Metal carboxylate conformation types. Magnetic behavior of
this class of materials strongly depends on the conformation type.
$M$ means metal and $L$ stands for ligand.} \label{conformations}
\end{figure}

\begin{figure*}[th]
\begin{center}
\subfigure[] {\includegraphics[width=3.0cm]{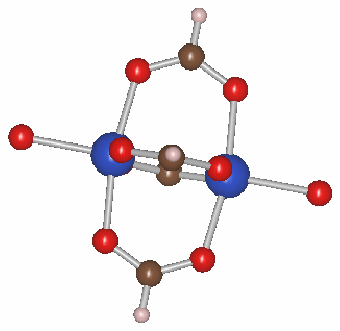}}\hfill
\subfigure[] {\includegraphics[width=3.0cm]{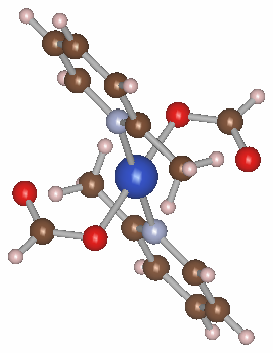}}\hfill
\subfigure[] {\includegraphics[width=9cm]{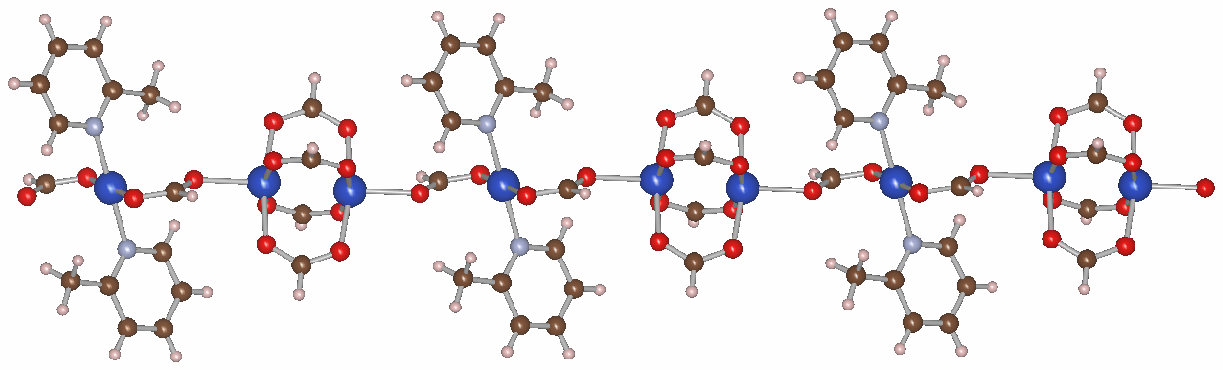}}
\end{center}\caption{(Color on line) Detailed view of the structural motifs of
\{Cu2(O2CH)4\}\{Cu(O2CH)2(2-methylpyridine)2\}. a) Dicopper
tetracarboxylate dimer unit. b) Copper dimethyl-pyridine monomeric
unit and c) view of the chain formed by alternating dimers and
monomers.}\label{estrutura1}
\end{figure*}

Metal carboxylates \cite{rao,sesto,blundell,perez} are
compounds that can present a
wide variety of topologies, compositions, and also allow 
multiple conformation environments, as shown in Fig.\ref{conformations}. A particularly
interesting case of conformation in these compounds is the syn-syn. In this
structure, a metal poli-carboxylate cluster, usually
called paddle-wheel (see figure \ref{estrutura1}(a)), is formed.
These clusters are  characterized by a four bridged M-M unit (i.e.,
a dimer), where the M ions are in square pyramidal coordination with
parallel basal planes. The available superexchange pathways observed
in these compounds, instead of direct exchange, allows an antiferromagnetic (AF) magnetic
exchange ($J/k_B$) of magnitude in the order of hundreds of degrees. This can be understood by recognizing that the
unpaired electron occupies the dx$^2$-y$^2$ orbital pointing to the
bridging oxygen, while the overlap between dz$^2$ orbitals is small
\cite{fortea}. The magnetic interaction within the paddle-wheel
dimer is therefore consistently both strong and antiferromagnetic, features that 
allows high entanglement temperatures \cite{pinto}. Despite their strong intra-dimer interaction, these paddle-wheel
compounds may still retain their low dimensional character due to
the large distances between the magnetic centers. Conversely, the
syn-anti and anti-anti conformations exhibits a rather weak magnetic
interaction, that can be either ferromagnetic (FM) or
antiferromagnetic, depending mainly on the nature of the ligand
(L) and the planarity of the carboxylate group.

Therefore, we have chosen a specific compound, namely
\{Cu$_2$(O$_2$CH)$_4$\}\{Cu(O$_2$CH)$_2$(2-methylpyridine)$_2$\},
that consists of copper dimer and copper monomers. The dimer is
formed by opposing square pyramidal CuO$_5$. The base oxygens on the
adjoining pyramids are part of the four-connecting carboxylate
groups in a syn-syn conformation which leads to a strong magnetic
interaction between the dimer's atoms. The Cu-Cu distance in this
ensemble is 2.63 \AA. This square pyramid is slightly distorted form the tetrahedral shape, with an average O$_{ap}-$Cu$-$O$_{eq}$ angle of about
95$^o$. The apical oxygen of the pyramid is connected, via another
carboxylate group, to the Cu(2-methylpyridine) in a syn-anti
configuration. This copper, the monomer, is in a pseudo-octahedral
coordination with four oxygens - two from each carboxylate group
along the chain and two opposing nitrogen ions from the
methyl-pyridine group. These alternating dimers and
monomers extend in one direction forming a chain of alternating
dimers and monomers (or a syn-syn-anti progression) where the
magnetic interaction between the dimer's atoms is strong, and weak
between the dimer and the monomer. The 2-methylpyridine groups
coordinating the monomeric copper ion act as spacers which are
placed alternating along the chain. This
large methylpyridine group, as well as the absence of any exchange
path between chains, prevents any significant inter-chain magnetic
interaction, making this system magnetically one-dimensional down to
the lowest measured temperature. The Table (\ref{table_bonds}) lists
some selected structural parameters for this compound.
\begin{table}
  \begin{center}
  \begin{tabular}{|c|c||c|c|}
    \hline
Cu$_d-$Cu$_d$ &  2.631 \AA  & Cu$-$O$_m$ & 1.976 \AA \\\hline
Cu$_d-$Cu$_m$ & 4.689 \AA & O$-$C$-$O$_d$ & 127.9$^o$\\\hline
Cu$_d-$O$_{ap}$ & 2.120$^o$ & O$-$C$-$O$_m$ & 123.6$^o$ \\\hline
$\langle$ Cu$_d-$O$_{eq}$ $\rangle$ &  1.978 \AA & Cu$-$Cu
(interchain) & 8.1055 \AA \\\hline
Cu$_d-$O$-$Cu$_m$ & 160$^o$& &
\\\hline    \hline
  \end{tabular}
  \end{center}
  \caption{Selected bond lengths and angles, relevant for the magnetic properties observed  for compound
  \{Cu$_2$(O$_2$CH)$_4$\}\{Cu(O$_2$CH)$_2$(2-methylpyridine)$_2$\}. Cu$_d$ - dimer copper, Cu$_m$ - monomer copper, O$_{ap}$ - apical oxigen, O$_{eq}$ - equatorial oxigen. }
  \label{table_bonds}
\end{table}

\begin{figure}[tbp]
\begin{center}
\includegraphics[width=8.0cm]{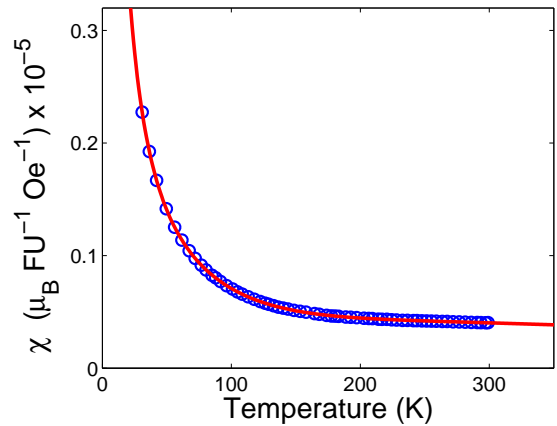}
\end{center}
\caption{Magnetic susceptibility as a function of temperature
with an applied field of $100$ Oe. The
points are the experimental results and the solid line is the
theoretical prediction, based on the dimer-monomer model, as
discussed in the text.} \label{x}
\end{figure}

Since the syn-anti magnetic
interaction is typically weaker than the syn-syn
\cite{colacio,zurowska}, the magnetic properties of the compound can be 
modeled considering a superposition of a dimer susceptibility with a Curie-Weiss
type susceptibility:
\begin{equation}
  \chi = \chi_d +  \chi_m.
\label{chin} 
\end{equation}
The first term
corresponds to the dimer magnetic susceptibility and for low
magnetic fields it is given by \cite{khan}:
\begin{equation}
\chi_d = \frac{(g \mu_B )^2}{k_B T} \frac{2}{3 + e^{-J/k_B T}},
\end{equation}
where $g$ is the Land� factor, $\mu_B$ is the Bohr magneton and $k_B$ is the 
Boltzmann constant. The second term in (\ref{chin}) represents the magnetic susceptibility of the
monomer and,  since it only interacts with a static magnetic
field, its susceptibility just follows the Curie law $\chi_m = C/T$. The experimental 
results were then fitted
according to this model and the parameters were found to be
$J/k_B = -693.15$ K, g = $2.21$ and $C = 7.02 \times 10^{-5}$ K
$\mu_B$ FU$^{-1}$ Oe$^{-1}$. In figure (\ref{x}) we show a comparison between 
the model and experimental data.

\section{Entanglement \label{sec3}}
 
\begin{figure}[tbp]
\begin{center}
\includegraphics[width=8.0cm]{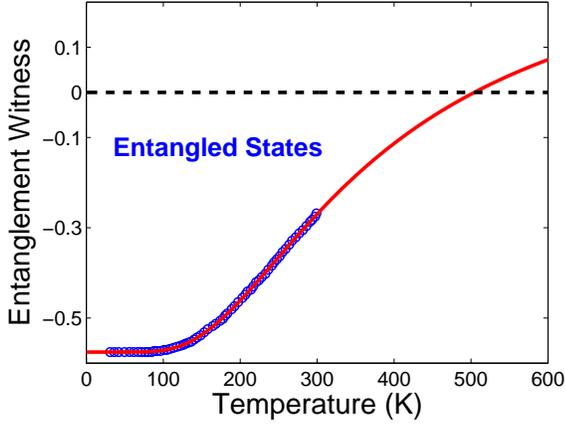}
\end{center}
\caption{Experimental entanglement
witness derived from the magnetic susceptibility measurements. The
points are the experimental results and the solid line is the
theoretical prediction, based on the dimer-monomer model, as
discussed in the text.} \label{ew}
\end{figure}

The task of entanglement quantification is still an open problem 
in general case (for a recent review see \cite{horodeckirev}). Hence, usually the detection of entanglement is done 
using a quantity called {\em Entanglement Witness} (EW). The concept of Entanglement Witness was first introduced 
by Horodecki et al. \cite{Horodecki}. An EW 
is an observable which is capable to identify whether a system is in an 
entangled state. For a spin chain, such a witness can be 
directly proportional to the magnetic susceptibility
\cite{wiesniak}:
\begin{equation}
EW(N)  = \frac{3 k_B T \bar{\chi}(T)}{(g\mu_B)^2 N S} - 1, \label{wit}
\end{equation}
where $N$ is the number of spin$-S$ particles and $\bar{\chi}$ is
the  average of the magnetic susceptibility measured along the three
ortogonal directions. For this witness, there will be entanglement in
the system if $EW(N) < 0$.  In figure \ref{ew}, the entanglement
witness obtained from the measured magnetic susceptibility is shown
as a function of the temperature. The witness is
negative up to nearly room temperature, showing the presence of
entanglement in the system.

To quantify the amount 
of entanglement in the dimer, we can use a
quantity called {\em Concurrence} \cite{wootters}. For two qubits
described by the quantum state $\rho$, the concurrence $\mathcal{C}$ is defined as \cite{wootters}:
\begin{equation}
\mathcal{C} = \max(0, \sqrt{\Lambda_1} -  \sqrt{\Lambda_2} -
\sqrt{\Lambda_3}  - \sqrt{\Lambda_4}),
\end{equation}
where $\Lambda$'s are the eigenvalues of $R = \rho \sigma_y \otimes
\sigma_y \rho^* \sigma_y \otimes \sigma_y$, labeled in decreasing
order. The degree of  entanglement, obtained form this quantity,
varies from $0$ to $1$, and a pair of spins is considered to be in a
maximally entangled state if $\mathcal{C} = 1$ and separable when
$\mathcal{C} = 0$.  For any other values the state of the spins is
said to be partially entangled. Using the dimer density matrix
$\rho_d$, it is possible to show that:
\begin{eqnarray}
\mathcal{C} &=& \max \left [ 0, - \frac{6}{3 + e^{-J/k_B T}} +1 \right ]    \label{c} \\
  &=& \max \left [ 0, - \frac{3 k_B T(\chi - C/T)}{(g \mu_B)^2} +1 \right ]
  \label{cc}
\end{eqnarray}

\begin{figure}[tbp]
\begin{center}
\includegraphics[width=8.0cm]{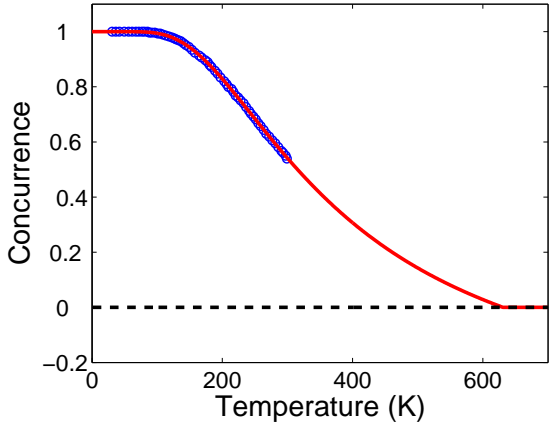}
\end{center}
\caption{Concurrence as a function of temperature. The points
are the experimental results and the solid line is the theoretical
prediction, based on Eq. (\ref{c}). } \label{cfig}
\end{figure}

The equation (\ref{cc}) shows that  the concurrence of the
dimer is also related to the magnetic susceptibility, which can be
obtained experimentally. On figure \ref{cfig} the concurrence
calculated according to the equation (\ref{cc}) is shown, and
the solid line is the theoretical prediction of the equation
(\ref{c}) with parameters $g$, $J$, and $C$ obtained from the
fit to the experimental susceptibility (see Fig. \ref{ew}). An interesting
result is
that the spins of the dimer remain maximally entangled up to $\sim
100$ K. From equation (\ref{c}), we can estimate the critical temperature
below which entanglement exists as being $T_e = -J/k_B \ln(3) \sim
630 $ K, which is well above room temperature. It is important to emphasize that the
high value of the exchange integral $J$ is due to the syn-syn
conformation and thus any material with such kind of conformation 
are strong candidates to contain
entanglement at high temperatures.

\begin{figure}[tbp]
\begin{center}
\includegraphics[width=8.0cm]{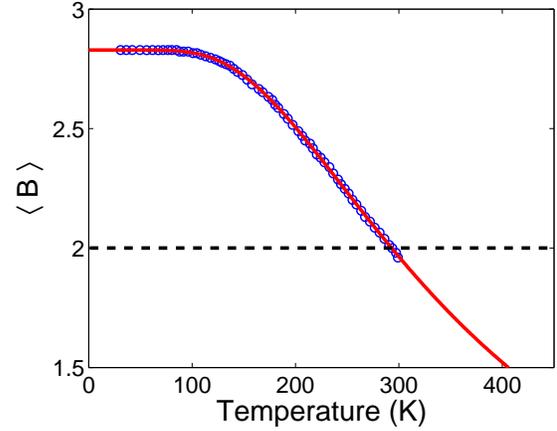}
\end{center}
\caption{The mean value of the Bell
operator as a function of temperature. The points are the
experimental results and the solid line is the theoretical
prediction, based on Eq. (\ref{bb}).} \label{bfig}
\end{figure}

The violation of Bell's inequalities is of great importance to quantum information science. These
inequalities are closely related to the
usefulness of entangled states \cite{brukner2,acin}. In particular,
Bell's inequalities violations are related to the security of
cryptography protocols \cite{acin} and are a necessary and
sufficient condition to the usefulness of quantum states in
communication complexity protocols \cite{brukner2}.  A Bell's
inequality test for two qubits involves measurements of a set of
correlations functions, which for a magnetic system are the
correlations of the magnetic moments along specific directions \cite{souzanjp}. For
a system with two spin $1/2$, the test involves the measurement of the
mean value of the Bell operator given by:
\begin{equation}
\mathcal{B} = \vec{n}_1 \cdot \vec{\sigma} \otimes (\vec{n}_2 \cdot \vec{\sigma} - \vec{n}_4 \cdot \vec{\sigma}) +
\vec{n}_3 \cdot \vec{\sigma} \otimes (\vec{n}_2 \cdot \vec{\sigma} + \vec{n}_4 \cdot \vec{\sigma})
\label{eqb}
\end{equation}

In the above equation, $\vec{n} \cdot \vec{\sigma}$ is  the
projection of the spin in the direction $\vec{n}$. For any separable
state, the mean value of Eq. (\ref{eqb}) satisfies the relation $|
\langle \mathcal{B} \rangle| \leq +2$, and whenever this inequality
is violated, the system is in an entangled state \cite{genovese}.
There is a particular set of directions for which the violation
reaches its maximum. The ground state of an antiferromagnetic dimer
violates
maximally the Bell's inequality if we choose $\vec{n}_{1}$, $%
\vec{n}_{2}$, $\vec{n}_{3}$ and $\vec{n}_{4}$ as being  $(0,0,-1)$%
, $(-1,0,-1)/\sqrt{2}$, $(-1,0,0)$ and $(-1,0,1)/\sqrt{2}$,
respectively.  Then, using this set of directions, the Bell operator becomes:
 $\mathcal{B} = \sqrt{2} (\sigma_z \otimes \sigma_z + \sigma_x \otimes \sigma_x)$. Using
the dimer density matrix, $\rho_d$, to calculate the
correlations $\langle \sigma_z \otimes \sigma_z \rangle$ and $\langle \sigma_x \otimes \sigma_x \rangle$,  it
is easy to show that:
\begin{eqnarray}
|\langle \mathcal{B} \rangle | &=& 4 \sqrt{2} \left| \frac{2}{3 + e^{-J/k_B T}} -\frac{1}{2} \right |  \label{bb} \\
                  &=& 4 \sqrt{2}  \left| \frac{k_B T (\chi - C/T)}{(g \mu_B )^2} -\frac{1}{2} \right |
                  \label{bm}
\end{eqnarray}

From equation (\ref{bm}), it is possible to verify whether the entangled state of the
system violates the Bell's inequality or not. On figure \ref{bfig}, the mean value of the
Bell operator as a function of the temperature is shown, along with
its theoretical prediction. From the figure
\ref{bfig}, we see that the Bell's inequality is violated below $\sim 290$ K
and maximum violation is observed for temperatures below $ \sim 100$ K, which is
compatible to the previous result, obtained from the concurrence,
i.e., the system remains maximally entangled up to $\sim 100$ K).

\section{Conclusion}

In summary, we identified a family of magnetic materials
that can support entanglement at very high temperatures, namely
metal carboxylates with syn-syn conformation. As an example, we choose the compound
\{Cu$_2$(O$_2$CH)$_4$\}\{Cu(O$_2$CH)$_2$(2-methylpyridine)$_2$\},
and for this material we determine that the critical temperature
below bipartite entanglement exist is $T_e  \sim 630$ K.
Furthermore, we could also conclude that the system remains
maximally entangled up to $\sim 100$ K and the Bell's inequality can
be violated up to close room temperature ($ \sim 290$ K), which is a special
feature of this material, since Bell's inequalities
are of great importance to Quantum Computation \cite{acin,brukner2}.  The
critical temperature estimated in this work is the highest value
reported in the literature and the identification of this
class of materials can open the way doors for new researches toward a solid
state quantum devices. In addition, we are convinced that other systems
can support entanglement at even higher temperatures, since exchange
magnetic coupling as high as that provided here has been reported in
the literature \cite{motoyama}.

\begin{acknowledgments}
The authors acknowledge support from the Brazilian funding agencies
CNPq,  CAPES and the Brazilian Millennium Institute for Quantum
Information. MSR thanks the financial support from PCI-CBPF program.
Research partially sponsored by the Laboratory Directed Research and
Development Program and the Division of Materials Sciences and
Engineering, of Oak Ridge National Laboratory (ORNL), managed by
UT-Battelle, LLC for the U. S. Department of Energy under Contract
No. DE-AC05-00OR22725.
\end{acknowledgments}

\bibliography{DimMonCu}

\begin{thebibliography}{38}
\expandafter\ifx\csname natexlab\endcsname\relax\def\natexlab#1{#1}\fi
\expandafter\ifx\csname bibnamefont\endcsname\relax
  \def\bibnamefont#1{#1}\fi
\expandafter\ifx\csname bibfnamefont\endcsname\relax
  \def\bibfnamefont#1{#1}\fi
\expandafter\ifx\csname citenamefont\endcsname\relax
  \def\citenamefont#1{#1}\fi
\expandafter\ifx\csname url\endcsname\relax
  \def\url#1{\texttt{#1}}\fi
\expandafter\ifx\csname urlprefix\endcsname\relax\def\urlprefix{URL }\fi
\providecommand{\bibinfo}[2]{#2}
\providecommand{\eprint}[2][]{\url{#2}}

\bibitem[{\citenamefont{Nielsen and Chuang}(2000)}]{nielsenB}
\bibinfo{author}{\bibfnamefont{M.~A.} \bibnamefont{Nielsen}} \bibnamefont{and}
  \bibinfo{author}{\bibfnamefont{I.~L.} \bibnamefont{Chuang}},
  \emph{\bibinfo{title}{Quantum Computation and Quantum Information}}
  (\bibinfo{publisher}{Cambridge University Press},
  \bibinfo{address}{Cambridge}, \bibinfo{year}{2000}).

\bibitem[{\citenamefont{Oliveira et~al.}(2007)\citenamefont{Oliveira,
  Bonagamba, Sarthour, Freitas, and deAzevedo}}]{oliveira}
\bibinfo{author}{\bibfnamefont{I.~S.} \bibnamefont{Oliveira}},
  \bibinfo{author}{\bibfnamefont{T.~J.} \bibnamefont{Bonagamba}},
  \bibinfo{author}{\bibfnamefont{R.~S.} \bibnamefont{Sarthour}},
  \bibinfo{author}{\bibfnamefont{J.~C.~C.} \bibnamefont{Freitas}},
  \bibnamefont{and} \bibinfo{author}{\bibfnamefont{E.~R.}
  \bibnamefont{deAzevedo}}, \emph{\bibinfo{title}{NMR Quantum Information
  Processing}} (\bibinfo{publisher}{Elsevier}, \bibinfo{address}{Copenhagen,
  Neatherland}, \bibinfo{year}{2007}).

\bibitem[{\citenamefont{Arnesen et~al.}(2001)\citenamefont{Arnesen, Bose, and
  Vedral}}]{vedralprl}
\bibinfo{author}{\bibfnamefont{M.~C.} \bibnamefont{Arnesen}},
  \bibinfo{author}{\bibfnamefont{S.}~\bibnamefont{Bose}}, \bibnamefont{and}
  \bibinfo{author}{\bibfnamefont{V.}~\bibnamefont{Vedral}},
  \bibinfo{journal}{Phys. Rev. Lett.} \textbf{\bibinfo{volume}{87}},
  \bibinfo{pages}{017901} (\bibinfo{year}{2001}).

\bibitem[{\citenamefont{Amico et~al.}(2008)\citenamefont{Amico, Fazio,
  Osterloh, and Vedral}}]{vedralrmp}
\bibinfo{author}{\bibfnamefont{L.}~\bibnamefont{Amico}},
  \bibinfo{author}{\bibfnamefont{R.}~\bibnamefont{Fazio}},
  \bibinfo{author}{\bibfnamefont{A.}~\bibnamefont{Osterloh}}, \bibnamefont{and}
  \bibinfo{author}{\bibfnamefont{V.}~\bibnamefont{Vedral}},
  \bibinfo{journal}{Rev. Mod. Phys.} \textbf{\bibinfo{volume}{80}},
  \bibinfo{pages}{517} (\bibinfo{year}{2008}).

\bibitem[{\citenamefont{Markham et~al.}(2008)\citenamefont{Markham, Anders,
  Vedral, Murao, and Miyake}}]{vedralepl}
\bibinfo{author}{\bibfnamefont{D.}~\bibnamefont{Markham}},
  \bibinfo{author}{\bibfnamefont{J.}~\bibnamefont{Anders}},
  \bibinfo{author}{\bibfnamefont{V.}~\bibnamefont{Vedral}},
  \bibinfo{author}{\bibfnamefont{M.}~\bibnamefont{Murao}}, \bibnamefont{and}
  \bibinfo{author}{\bibfnamefont{A.}~\bibnamefont{Miyake}},
  \bibinfo{journal}{Europhys. Lett.} \textbf{\bibinfo{volume}{81}},
  \bibinfo{pages}{40006} (\bibinfo{year}{2008}).

\bibitem[{\citenamefont{Ghosh et~al.}(2003)\citenamefont{Ghosh, Rosenbaum,
  Aeppli, and Coppersmith}}]{ghosh}
\bibinfo{author}{\bibfnamefont{S.}~\bibnamefont{Ghosh}},
  \bibinfo{author}{\bibfnamefont{T.~F.} \bibnamefont{Rosenbaum}},
  \bibinfo{author}{\bibfnamefont{G.}~\bibnamefont{Aeppli}}, \bibnamefont{and}
  \bibinfo{author}{\bibfnamefont{S.~N.} \bibnamefont{Coppersmith}},
  \bibinfo{journal}{Nature} \textbf{\bibinfo{volume}{452}}, \bibinfo{pages}{48}
  (\bibinfo{year}{2003}).

\bibitem[{\citenamefont{\v{C}. Brukner et~al.}(2006)\citenamefont{\v{C}.
  Brukner, Vedral, and Zeilinger}}]{brukner}
\bibinfo{author}{\bibnamefont{\v{C}. Brukner}},
  \bibinfo{author}{\bibfnamefont{V.}~\bibnamefont{Vedral}}, \bibnamefont{and}
  \bibinfo{author}{\bibfnamefont{A.}~\bibnamefont{Zeilinger}},
  \bibinfo{journal}{Phys. Rev. A} \textbf{\bibinfo{volume}{73}},
  \bibinfo{pages}{012110} (\bibinfo{year}{2006}).

\bibitem[{\citenamefont{Rappoport et~al.}(2007)\citenamefont{Rappoport,
  Ghivelder, Fernandes, Guimar{\~a}es, and Continentino}}]{rappoport}
\bibinfo{author}{\bibfnamefont{T.~G.} \bibnamefont{Rappoport}},
  \bibinfo{author}{\bibfnamefont{L.}~\bibnamefont{Ghivelder}},
  \bibinfo{author}{\bibfnamefont{J.~C.} \bibnamefont{Fernandes}},
  \bibinfo{author}{\bibfnamefont{R.~B.} \bibnamefont{Guimar{\~a}es}},
  \bibnamefont{and} \bibinfo{author}{\bibfnamefont{M.~A.}
  \bibnamefont{Continentino}}, \bibinfo{journal}{Phys. Rev. B}
  \textbf{\bibinfo{volume}{75}}, \bibinfo{pages}{054422}
  (\bibinfo{year}{2007}).

\bibitem[{\citenamefont{V\'ertesi and Bene}(2006)}]{vertesi}
\bibinfo{author}{\bibfnamefont{T.}~\bibnamefont{V\'ertesi}} \bibnamefont{and}
  \bibinfo{author}{\bibfnamefont{E.}~\bibnamefont{Bene}},
  \bibinfo{journal}{Phys. Rev. B} \textbf{\bibinfo{volume}{73}},
  \bibinfo{pages}{134404} (\bibinfo{year}{2006}).

\bibitem[{\citenamefont{Souza et~al.}(2008{\natexlab{a}})\citenamefont{Souza,
  Reis, Soares-Pinto, Oliveira, and Sarthour}}]{souza}
\bibinfo{author}{\bibfnamefont{A.~M.} \bibnamefont{Souza}},
  \bibinfo{author}{\bibfnamefont{M.~S.} \bibnamefont{Reis}},
  \bibinfo{author}{\bibfnamefont{D.~O.} \bibnamefont{Soares-Pinto}},
  \bibinfo{author}{\bibfnamefont{I.~S.} \bibnamefont{Oliveira}},
  \bibnamefont{and} \bibinfo{author}{\bibfnamefont{R.~S.}
  \bibnamefont{Sarthour}}, \bibinfo{journal}{Phys. Rev. B}
  \textbf{\bibinfo{volume}{77}}, \bibinfo{pages}{104402}
  (\bibinfo{year}{2008}{\natexlab{a}}).

\bibitem[{\citenamefont{Soares-Pinto et~al.}(2008)\citenamefont{Soares-Pinto,
  Souza, Sarthour, Oliveira, Reis, ao, and dos Santos}}]{pinto}
\bibinfo{author}{\bibfnamefont{D.~O.} \bibnamefont{Soares-Pinto}},
  \bibinfo{author}{\bibfnamefont{A.~M.} \bibnamefont{Souza}},
  \bibinfo{author}{\bibfnamefont{R.~S.} \bibnamefont{Sarthour}},
  \bibinfo{author}{\bibfnamefont{I.~S.} \bibnamefont{Oliveira}},
  \bibinfo{author}{\bibfnamefont{M.~S.} \bibnamefont{Reis}},
  \bibinfo{author}{\bibfnamefont{P.~B.} \bibnamefont{ao}}, \bibnamefont{and}
  \bibinfo{author}{\bibfnamefont{A.~M.} \bibnamefont{dos Santos}},
  \bibinfo{journal}{Submitted to New J. Phys.}  (\bibinfo{year}{2008}).

\bibitem[{\citenamefont{Vedral}(2008)}]{vedral}
\bibinfo{author}{\bibfnamefont{V.}~\bibnamefont{Vedral}},
  \bibinfo{journal}{Nature} \textbf{\bibinfo{volume}{453}},
  \bibinfo{pages}{1004} (\bibinfo{year}{2008}).

\bibitem[{\citenamefont{Chiara et~al.}(2006)\citenamefont{Chiara, \v{C}.
  Brukner R.~Fazio, Palma, and Vedral}}]{chiara06}
\bibinfo{author}{\bibfnamefont{G.~D.} \bibnamefont{Chiara}},
  \bibinfo{author}{\bibnamefont{\v{C}. Brukner R.~Fazio}},
  \bibinfo{author}{\bibfnamefont{G.~M.} \bibnamefont{Palma}}, \bibnamefont{and}
  \bibinfo{author}{\bibfnamefont{V.}~\bibnamefont{Vedral}},
  \bibinfo{journal}{New J. Phys.} \textbf{\bibinfo{volume}{8}},
  \bibinfo{pages}{95} (\bibinfo{year}{2006}).

\bibitem[{\citenamefont{\v{C}. Brukner et~al.}(2004)\citenamefont{\v{C}.
  Brukner, \.{Z}ukowski, Pan, and Zeilinger}}]{brukner2}
\bibinfo{author}{\bibnamefont{\v{C}. Brukner}},
  \bibinfo{author}{\bibfnamefont{M.}~\bibnamefont{\.{Z}ukowski}},
  \bibinfo{author}{\bibfnamefont{J.~W.} \bibnamefont{Pan}}, \bibnamefont{and}
  \bibinfo{author}{\bibfnamefont{A.}~\bibnamefont{Zeilinger}},
  \bibinfo{journal}{Phys. Rev. Lett.} \textbf{\bibinfo{volume}{92}},
  \bibinfo{pages}{127901} (\bibinfo{year}{2004}).

\bibitem[{\citenamefont{Acín et~al.}(2006)\citenamefont{Acín, Gisin, and
  Masanes}}]{acin}
\bibinfo{author}{\bibfnamefont{A.}~\bibnamefont{Acín}},
  \bibinfo{author}{\bibfnamefont{N.}~\bibnamefont{Gisin}}, \bibnamefont{and}
  \bibinfo{author}{\bibfnamefont{L.}~\bibnamefont{Masanes}},
  \bibinfo{journal}{Phys, Rev. Lett.} \textbf{\bibinfo{volume}{97}},
  \bibinfo{pages}{120405} (\bibinfo{year}{2006}).

\bibitem[{\citenamefont{Bogani and Wernsdorfer}(2008)}]{bogani}
\bibinfo{author}{\bibfnamefont{L.}~\bibnamefont{Bogani}} \bibnamefont{and}
  \bibinfo{author}{\bibfnamefont{W.}~\bibnamefont{Wernsdorfer}},
  \bibinfo{journal}{Nature Materials} \textbf{\bibinfo{volume}{7}},
  \bibinfo{pages}{179} (\bibinfo{year}{2008}).

\bibitem[{\citenamefont{Ramsay et~al.}(2008)\citenamefont{Ramsay, del Barco,
  Hill, Shah, Beedle, and Hendrickson}}]{ramsay}
\bibinfo{author}{\bibfnamefont{C.~M.} \bibnamefont{Ramsay}},
  \bibinfo{author}{\bibfnamefont{E.}~\bibnamefont{del Barco}},
  \bibinfo{author}{\bibfnamefont{S.}~\bibnamefont{Hill}},
  \bibinfo{author}{\bibfnamefont{S.~J.} \bibnamefont{Shah}},
  \bibinfo{author}{\bibfnamefont{C.~C.} \bibnamefont{Beedle}},
  \bibnamefont{and} \bibinfo{author}{\bibfnamefont{D.~N.}
  \bibnamefont{Hendrickson}}, \bibinfo{journal}{Nature Physics}
  \textbf{\bibinfo{volume}{4}}, \bibinfo{pages}{277} (\bibinfo{year}{2008}).

\bibitem[{\citenamefont{Ardavan et~al.}(2007)\citenamefont{Ardavan, Rival,
  Morton, Blundell, Tyryshkin, Timco, and Winpenny}}]{ardavan}
\bibinfo{author}{\bibfnamefont{A.}~\bibnamefont{Ardavan}},
  \bibinfo{author}{\bibfnamefont{O.}~\bibnamefont{Rival}},
  \bibinfo{author}{\bibfnamefont{J.~J.~L.} \bibnamefont{Morton}},
  \bibinfo{author}{\bibfnamefont{S.~J.} \bibnamefont{Blundell}},
  \bibinfo{author}{\bibfnamefont{A.~M.} \bibnamefont{Tyryshkin}},
  \bibinfo{author}{\bibfnamefont{A.}~\bibnamefont{Timco}}, \bibnamefont{and}
  \bibinfo{author}{\bibfnamefont{R.~E.~P.} \bibnamefont{Winpenny}},
  \bibinfo{journal}{Phys. Rev. Lett.} \textbf{\bibinfo{volume}{98}},
  \bibinfo{pages}{057201} (\bibinfo{year}{2007}).

\bibitem[{\citenamefont{Affronte et~al.}(2005)\citenamefont{Affronte, Casson,
  Evangelistia, Candini, Carretta, Muryna, Teat, Timcoa, Wernsdorfer, and
  Winpenny}}]{affronte}
\bibinfo{author}{\bibfnamefont{M.}~\bibnamefont{Affronte}},
  \bibinfo{author}{\bibfnamefont{I.}~\bibnamefont{Casson}},
  \bibinfo{author}{\bibfnamefont{M.}~\bibnamefont{Evangelistia}},
  \bibinfo{author}{\bibfnamefont{A.}~\bibnamefont{Candini}},
  \bibinfo{author}{\bibfnamefont{S.}~\bibnamefont{Carretta}},
  \bibinfo{author}{\bibfnamefont{C.}~\bibnamefont{Muryna}},
  \bibinfo{author}{\bibfnamefont{S.}~\bibnamefont{Teat}},
  \bibinfo{author}{\bibfnamefont{G.}~\bibnamefont{Timcoa}},
  \bibinfo{author}{\bibfnamefont{W.}~\bibnamefont{Wernsdorfer}},
  \bibnamefont{and} \bibinfo{author}{\bibfnamefont{R.}~\bibnamefont{Winpenny}},
  \bibinfo{journal}{Angew. Chem. Int} \textbf{\bibinfo{volume}{44}},
  \bibinfo{pages}{6496} (\bibinfo{year}{2005}).

\bibitem[{\citenamefont{Troiani
  et~al.}(2005{\natexlab{a}})\citenamefont{Troiani, Affronte, Carretta,
  Santini, and Amoretti}}]{troiani1}
\bibinfo{author}{\bibfnamefont{F.}~\bibnamefont{Troiani}},
  \bibinfo{author}{\bibfnamefont{M.}~\bibnamefont{Affronte}},
  \bibinfo{author}{\bibfnamefont{S.}~\bibnamefont{Carretta}},
  \bibinfo{author}{\bibfnamefont{P.}~\bibnamefont{Santini}}, \bibnamefont{and}
  \bibinfo{author}{\bibfnamefont{G.}~\bibnamefont{Amoretti}},
  \bibinfo{journal}{Phys. Rev. Lett} \textbf{\bibinfo{volume}{94}},
  \bibinfo{pages}{190501} (\bibinfo{year}{2005}{\natexlab{a}}).

\bibitem[{\citenamefont{Troiani
  et~al.}(2005{\natexlab{b}})\citenamefont{Troiani, Ghirri, Affronte, Carretta,
  Santini, Amoretti, Piligkos, Timco, and Winpenny}}]{troiani2}
\bibinfo{author}{\bibfnamefont{F.}~\bibnamefont{Troiani}},
  \bibinfo{author}{\bibfnamefont{A.}~\bibnamefont{Ghirri}},
  \bibinfo{author}{\bibfnamefont{M.}~\bibnamefont{Affronte}},
  \bibinfo{author}{\bibfnamefont{S.}~\bibnamefont{Carretta}},
  \bibinfo{author}{\bibfnamefont{P.}~\bibnamefont{Santini}},
  \bibinfo{author}{\bibfnamefont{G.}~\bibnamefont{Amoretti}},
  \bibinfo{author}{\bibfnamefont{S.}~\bibnamefont{Piligkos}},
  \bibinfo{author}{\bibfnamefont{G.}~\bibnamefont{Timco}}, \bibnamefont{and}
  \bibinfo{author}{\bibfnamefont{R.~E.~P.} \bibnamefont{Winpenny}},
  \bibinfo{journal}{Phys. Rev. Lett} \textbf{\bibinfo{volume}{94}},
  \bibinfo{pages}{207208} (\bibinfo{year}{2005}{\natexlab{b}}).

\bibitem[{\citenamefont{Michael et~al.}(2001)\citenamefont{Michael,
  Leuenberger, and Loss}}]{michael}
\bibinfo{author}{\bibfnamefont{F.}~\bibnamefont{Michael}},
  \bibinfo{author}{\bibfnamefont{N.}~\bibnamefont{Leuenberger}},
  \bibnamefont{and} \bibinfo{author}{\bibfnamefont{D.}~\bibnamefont{Loss}},
  \bibinfo{journal}{Nature} \textbf{\bibinfo{volume}{410}},
  \bibinfo{pages}{789} (\bibinfo{year}{2001}).

\bibitem[{\citenamefont{Lehmann et~al.}(2007)\citenamefont{Lehmann, Gaita-Ario,
  Coronado, and Loss}}]{lehmann}
\bibinfo{author}{\bibfnamefont{J.}~\bibnamefont{Lehmann}},
  \bibinfo{author}{\bibfnamefont{A.}~\bibnamefont{Gaita-Ario}},
  \bibinfo{author}{\bibfnamefont{E.}~\bibnamefont{Coronado}}, \bibnamefont{and}
  \bibinfo{author}{\bibfnamefont{D.}~\bibnamefont{Loss}},
  \bibinfo{journal}{Nature Nanotechnology} \textbf{\bibinfo{volume}{2}},
  \bibinfo{pages}{312} (\bibinfo{year}{2007}).

\bibitem[{\citenamefont{Rao et~al.}(2004)\citenamefont{Rao, Natarajan, and
  Vaidhyanathan}}]{rao}
\bibinfo{author}{\bibfnamefont{C.~N.~R.} \bibnamefont{Rao}},
  \bibinfo{author}{\bibfnamefont{S.}~\bibnamefont{Natarajan}},
  \bibnamefont{and}
  \bibinfo{author}{\bibfnamefont{R.}~\bibnamefont{Vaidhyanathan}},
  \bibinfo{journal}{Angew. Chem. Int. Ed.} \textbf{\bibinfo{volume}{43}},
  \bibinfo{pages}{1466} (\bibinfo{year}{2004}).

\bibitem[{\citenamefont{Sesto et~al.}(2002)\citenamefont{Sesto, Deakin, and
  Miller}}]{sesto}
\bibinfo{author}{\bibfnamefont{R.~E.} \bibnamefont{Sesto}},
  \bibinfo{author}{\bibfnamefont{L.}~\bibnamefont{Deakin}}, \bibnamefont{and}
  \bibinfo{author}{\bibfnamefont{J.~S.} \bibnamefont{Miller}},
  \bibinfo{journal}{Synthetic Metals} \textbf{\bibinfo{volume}{122}},
  \bibinfo{pages}{543} (\bibinfo{year}{2002}).

\bibitem[{\citenamefont{Blundell and Pratt}(2004)}]{blundell}
\bibinfo{author}{\bibfnamefont{S.~J.} \bibnamefont{Blundell}} \bibnamefont{and}
  \bibinfo{author}{\bibfnamefont{F.~L.} \bibnamefont{Pratt}},
  \bibinfo{journal}{J. Phys.: Condens. Matter} \textbf{\bibinfo{volume}{16}},
  \bibinfo{pages}{R771} (\bibinfo{year}{2004}).

\bibitem[{\citenamefont{Calvo-P{\'e}rez
  et~al.}(2006)\citenamefont{Calvo-P{\'e}rez, Vega, and Spodine}}]{perez}
\bibinfo{author}{\bibfnamefont{V.}~\bibnamefont{Calvo-P{\'e}rez}},
  \bibinfo{author}{\bibfnamefont{A.}~\bibnamefont{Vega}}, \bibnamefont{and}
  \bibinfo{author}{\bibfnamefont{E.}~\bibnamefont{Spodine}},
  \bibinfo{journal}{Organometallics} \textbf{\bibinfo{volume}{25}},
  \bibinfo{pages}{1953} (\bibinfo{year}{2006}).

\bibitem[{\citenamefont{Rodríguez-Fortea
  et~al.}(2001)\citenamefont{Rodríguez-Fortea, Alemany, Alvarez, and
  Ruiz}}]{fortea}
\bibinfo{author}{\bibfnamefont{A.}~\bibnamefont{Rodríguez-Fortea}},
  \bibinfo{author}{\bibfnamefont{P.}~\bibnamefont{Alemany}},
  \bibinfo{author}{\bibfnamefont{S.}~\bibnamefont{Alvarez}}, \bibnamefont{and}
  \bibinfo{author}{\bibfnamefont{E.}~\bibnamefont{Ruiz}},
  \bibinfo{journal}{Chem. Eur. J.} \textbf{\bibinfo{volume}{7}},
  \bibinfo{pages}{627} (\bibinfo{year}{2001}).

\bibitem[{\citenamefont{Colacio et~al.}(1992)\citenamefont{Colacio,
  Dominguez-Vera, Costes, Kivekas, Laurent, Ruiz, and Sundberg}}]{colacio}
\bibinfo{author}{\bibfnamefont{E.}~\bibnamefont{Colacio}},
  \bibinfo{author}{\bibfnamefont{J.~M.} \bibnamefont{Dominguez-Vera}},
  \bibinfo{author}{\bibfnamefont{J.~P.} \bibnamefont{Costes}},
  \bibinfo{author}{\bibfnamefont{R.}~\bibnamefont{Kivekas}},
  \bibinfo{author}{\bibfnamefont{J.~P.} \bibnamefont{Laurent}},
  \bibinfo{author}{\bibfnamefont{J.}~\bibnamefont{Ruiz}}, \bibnamefont{and}
  \bibinfo{author}{\bibfnamefont{M.}~\bibnamefont{Sundberg}},
  \bibinfo{journal}{Inorg. Chem.} \textbf{\bibinfo{volume}{31}},
  \bibinfo{pages}{774} (\bibinfo{year}{1992}).

\bibitem[{\citenamefont{{\. Z}urowska et~al.}(2007)\citenamefont{{\. Z}urowska,
  Mrozi{\' n}ski, and Ciunik}}]{zurowska}
\bibinfo{author}{\bibfnamefont{B.}~\bibnamefont{{\. Z}urowska}},
  \bibinfo{author}{\bibfnamefont{J.}~\bibnamefont{Mrozi{\' n}ski}},
  \bibnamefont{and} \bibinfo{author}{\bibfnamefont{Z.}~\bibnamefont{Ciunik}},
  \bibinfo{journal}{Polyhedron} \textbf{\bibinfo{volume}{26}},
  \bibinfo{pages}{3085} (\bibinfo{year}{2007}).

\bibitem[{\citenamefont{Khan}(1993)}]{khan}
\bibinfo{author}{\bibfnamefont{O.}~\bibnamefont{Khan}},
  \emph{\bibinfo{title}{Molecular Magnetism}} (\bibinfo{publisher}{Wiley-VCH},
  \bibinfo{address}{New York}, \bibinfo{year}{1993}).

\bibitem[{\citenamefont{Horodecki et~al.}(2008)\citenamefont{Horodecki,
  Horodecki, Horodecki, and Horodecki}}]{horodeckirev}
\bibinfo{author}{\bibfnamefont{R.}~\bibnamefont{Horodecki}},
  \bibinfo{author}{\bibfnamefont{P.}~\bibnamefont{Horodecki}},
  \bibinfo{author}{\bibfnamefont{M.}~\bibnamefont{Horodecki}},
  \bibnamefont{and}
  \bibinfo{author}{\bibfnamefont{K.}~\bibnamefont{Horodecki}},
  \bibinfo{journal}{Accepted in Rev. Mod. Phys.}  (\bibinfo{year}{2008}).

\bibitem[{\citenamefont{Horodecki et~al.}(1996)\citenamefont{Horodecki,
  Horodecki, and Horodecki}}]{Horodecki}
\bibinfo{author}{\bibfnamefont{M.}~\bibnamefont{Horodecki}},
  \bibinfo{author}{\bibfnamefont{P.}~\bibnamefont{Horodecki}},
  \bibnamefont{and}
  \bibinfo{author}{\bibfnamefont{R.}~\bibnamefont{Horodecki}},
  \bibinfo{journal}{Phys. Lett. A} \textbf{\bibinfo{volume}{223}},
  \bibinfo{pages}{1} (\bibinfo{year}{1996}).

\bibitem[{\citenamefont{Wie\'sniak et~al.}(2005)\citenamefont{Wie\'sniak,
  Vedral, and \v{C}. Brukner}}]{wiesniak}
\bibinfo{author}{\bibfnamefont{M.~.} \bibnamefont{Wie\'sniak}},
  \bibinfo{author}{\bibfnamefont{V.}~\bibnamefont{Vedral}}, \bibnamefont{and}
  \bibinfo{author}{\bibnamefont{\v{C}. Brukner}}, \bibinfo{journal}{New. J.
  Phys.} \textbf{\bibinfo{volume}{7}}, \bibinfo{pages}{258}
  (\bibinfo{year}{2005}).

\bibitem[{\citenamefont{Wootters}(1998)}]{wootters}
\bibinfo{author}{\bibfnamefont{W.~K.} \bibnamefont{Wootters}},
  \bibinfo{journal}{Phys. Rev. Lett.} \textbf{\bibinfo{volume}{80}},
  \bibinfo{pages}{2245} (\bibinfo{year}{1998}).

\bibitem[{\citenamefont{Souza et~al.}(2008{\natexlab{b}})\citenamefont{Souza,
  Magalhães, Teles, deAzevedo, Bonagamba, Oliveira, and Sarthour}}]{souzanjp}
\bibinfo{author}{\bibfnamefont{A.~M.} \bibnamefont{Souza}},
  \bibinfo{author}{\bibfnamefont{A.}~\bibnamefont{Magalhães}},
  \bibinfo{author}{\bibfnamefont{J.}~\bibnamefont{Teles}},
  \bibinfo{author}{\bibfnamefont{E.~R.} \bibnamefont{deAzevedo}},
  \bibinfo{author}{\bibfnamefont{T.~J.} \bibnamefont{Bonagamba}},
  \bibinfo{author}{\bibfnamefont{I.~S.} \bibnamefont{Oliveira}},
  \bibnamefont{and} \bibinfo{author}{\bibfnamefont{R.~S.}
  \bibnamefont{Sarthour}}, \bibinfo{journal}{New J. Phys.}
  \textbf{\bibinfo{volume}{10}}, \bibinfo{pages}{033020}
  (\bibinfo{year}{2008}{\natexlab{b}}).

\bibitem[{\citenamefont{Genovese}(2005)}]{genovese}
\bibinfo{author}{\bibfnamefont{M.}~\bibnamefont{Genovese}},
  \bibinfo{journal}{Phys. Rep.} \textbf{\bibinfo{volume}{413}},
  \bibinfo{pages}{319} (\bibinfo{year}{2005}).

\bibitem[{\citenamefont{Motoyama et~al.}(1996)\citenamefont{Motoyama, Eisaki,
  and Uchida}}]{motoyama}
\bibinfo{author}{\bibfnamefont{N.}~\bibnamefont{Motoyama}},
  \bibinfo{author}{\bibfnamefont{H.}~\bibnamefont{Eisaki}}, \bibnamefont{and}
  \bibinfo{author}{\bibfnamefont{S.}~\bibnamefont{Uchida}},
  \bibinfo{journal}{Phys. Rev. Lett.} \textbf{\bibinfo{volume}{76}},
  \bibinfo{pages}{3212} (\bibinfo{year}{1996}).

\end{thebibliography}

\end{document}